\documentclass[reprint,aps,amsmath,amssymb]{revtex4-1}

\usepackage{graphicx}
\usepackage{dcolumn}
\usepackage{bm}
\usepackage{hyperref}

\usepackage{amssymb,euscript,latexsym,amsmath}
\usepackage{graphicx}
\usepackage[dvips]{epsfig}
\usepackage{color}
\usepackage{epstopdf}
\usepackage{setspace}
\usepackage{subfigure} 

\newcommand{\abs}[1]{\left| #1 \right|} 

\newcommand{\conj}[1]{\overline{#1}}
\newcommand{\real}[1]{\mathrm{Re} \left[ #1 \right]}

\newcommand{\eee}[1]{\mathrm{e}^{ #1 }}
\newcommand{\ii}{\mathrm{i}} 

\begin{document}

\preprint{APS/123-QED}





\title{Flagella-induced transitions in the collective behavior of confined microswimmers}

\author{Alan Cheng Hou Tsang and  Eva Kanso} \thanks{corresponding author: kanso@usc.edu}
 \affiliation{Aerospace and Mechanical Engineering, University of Southern California, Los Angeles, CA 90089}
\date{\today}

\begin{abstract}
Bacteria exist in a free-swimming state or in a sessile biofilm state. The transition from free-swimming to sessile mode is characterized by changes in  gene expression which alter, among others, the mechanics of flagellar motility. 
In this paper, we propose an idealized physical model to investigate the effects of  flagellar activity on the hydrodynamic interactions among a population of microswimmers. We show that decreasing flagellar activity induces a hydrodynamically-triggered transition in  confined microswimmers from turbulent-like swimming  to aggregation and clustering.  These results suggest that the interplay between  flagellar activity and hydrodynamic interactions provides a physical mechanism for coordinating collective behaviors in confined bacteria, with potentially profound implications on biofilm initiation.
\begin{description}
\item[PACS numbers]
47.63.Gd,  87.18.Fx, 87.18.Hf, 05.65.+b
\end{description}
\end{abstract}

\maketitle

Collective motion emerges in a wide range of natural systems, from fish schools~\cite{couzin:jtb2002a,couzin:n2005a} to bacterial colonies~\cite{cisneros:ef2007a,sokolov:prl2007a,sokolov:pre2009a,darnton:bj2010a,zhang:pnas2010a,dunkel:prl2013a}, 
and is believed to play important roles in the functioning and survival of the group. 
For example, many species of bacteria cyclically transition from a free-swimming planktonic state into a sessile biofilm state in response to environmental conditions~\cite{miller:armb2001a,edmunds:jbr2013a}.  In their biofilm state, bacterial aggregates are difficult to eradicate, posing problems for medicine and industry~\cite{donlan:cid2001a, petrova:jb2012a}. 
Several of the molecular and genetic changes accompanying  biofilm formation have been deciphered, however
the specific mechanisms triggering the gene expressions that promote biofilm formation are less well understood. 
Transition to the biofilm state is characterized by a suppression of flagellar activity and thus flagella-driven motility. 
In some cases, flagellar activity during the back and forth transitions from free swimming to biofilm formation is regulated mechanically by a molecule that physically inhibits flagellar rotation in the biofilm state~\cite{zorraquino:jb2013a}. 

\begin{figure}
\includegraphics[width=0.5\textwidth]{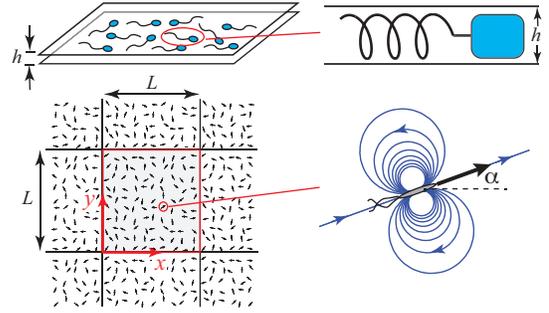}
\caption{\label{fig:results} Strongly confined micro-swimmers create potential dipolar far-field flows.}
\end{figure}

The question whether the interplay between flagellar activity and hydrodynamics could create a convective flow that lead to bacterial aggregation and biofilm initiation remains largely unexplored. In this paper, we show in the context of an idealized model that a decrease in flagellar activity could lead the microswimmers, via hydrodynamic interactions only, to transition from a turbulent-like swimming state, akin to the one observed in numerous experiments, to clustering and aggregation. 
The model considers asymmetric (head-tail) microswimmers  that are strongly confined in a thin film of Newtonian fluid, with the dimension of the swimmers being comparable to the thickness of the fluid film, see Figure~\ref{fig:results}. Geometric confinement is a common feature of the natural environment of various bacteria species. It is also characteristic of several experimental set-ups on biological and artificial microswimmers~\cite{darnton:bj2010a,zhang:pnas2010a,thutupalli:jp2011a, bricard:n2013a}.
Confined microswimmers have a distinct hydrodynamic signature 
in the sense that the far-field flow is that of a 2D \emph{potential source} dipole as opposed to the 3D \emph{force} dipole 
in the unbounded case~\cite{brotto:prl2013a}. Thus, the usual categorization of unbounded swimmers into pushers and pullers \cite{saintillan:prl2008a} becomes irrelevant. 
The dipolar far-field is independent of the transport mechanism (driven particles or self-propelled swimmers) 
and is rooted in the fact that the basic physics in confined fluids is that of a \emph{Hele-Shaw} potential flow~\cite{beatus:pr2012a}. 
Further,  due to friction with the nearby walls,  confined swimmers with geometric polarity (large head or large tail) reorient in response to both the local flow field and its gradient~\cite{brotto:prl2013a}, and their collective behavior exhibit instabilities that are qualitatively distinct from the ones observed in unbounded swimmers~\cite{brotto:prl2013a, lefauve:pre2014a}. 

In this paper, we seek a particle swimmer model that takes into account the effect  of flagellar activity. We first establish that the intensity of the  dipolar far-field induced by a beating flagellum depends on the level of flagellar activity: vigorously-beating flagella induce stronger dipolar fields than weakly-beating ones. The gap-averaged flagellar  motion is prescribed as $h(x,t)=A\cos(k x- t)$, with $x\in[-1,1]$, and is assumed to induce a constant swimming velocity $U$ in the $-x$-direction. Here, all parameters are dimensionless with the characteristic length and time scales being set by the flagellum's half-length and beating frequency, respectively.  The potential flow perturbation induced by this beating motion is computed numerically for various values of the beating amplitude $A$ and wavelength $k$ while normalizing the swimming velocity $U$ to 1. The time averaged flow field over one beating cycle is approximated, using a standard  fitting method based on minimization of  the $L^2$-norm, by the dipolar field of a circular disk with effective radius $R_{tail}$ moving at the same swimming velocity $U$, see Fig.~\ref{fig:flagella}. Note that the dipolar field induced by a circular disk located at $z_o = x_o + i y_o$ ($i = \sqrt{-1}$) in the complex $z$-plane and oriented at an arbitrary angle $\alpha_o$ to the $x$-axis can be described by the complex velocity $\conj{w}(z)=u_x-i u_y=\sigma \eee{\ii \alpha_o}/(z-z_o)^2$, where $\sigma = U R_{tail}^2$ is the dipole strength.
Fig.~\ref{fig:flagella}(c) shows that, as $A$ and $k$ increase, $R_{tail}$ increases accordingly. Consequently, the dipole strength $\sigma$ increases with increasing flagellar activity. 

We now model the flagellar far-field flow by that of a circular disk of effective radius $R_{tail}$ and we assume that the hydrodynamic-coupling between the head and the flagellum is weak. This leads to a head-tail dumbbell  swimmer model, where  the  value of $R_{tail}$ is interpreted as a measure of the flagellar activity. A derivation of the equations governing the motion of such weakly-coupled dumbbell swimmer can be found in~\cite{brotto:prl2013a, supp} and therefore is omitted here.
\begin{figure}
\includegraphics[width=0.45\textwidth]{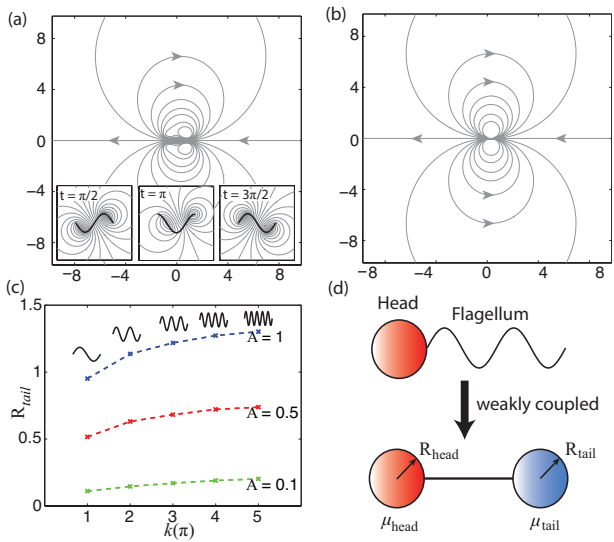}
\caption{\label{fig:flagella} (a) Time-averaged flow field created by a beating flagellum with $A=0.5$, $k=\pi$ and $U=1$. (Inset) Snapshots of the flagellum-induced flow field at different times. (b)  Dipolar field created by a circular disk of effective radius $R_{tail}$ fitted to the average flow field in (a). (c) Change in $R_{tail}$ with $A$ and $k$ of the traveling wave via the flagellum. (d) Reduction of a flagellated swimmer to a dumbbell swimmer.
\label{fig:flagella}}
\vskip -0.25in
\end{figure}
The dynamics of a population of $N$ such swimmers  can be expressed in concise complex notation
\begin{equation}
\begin{split}
\label{eq:formulation:eom}
  \dot{\conj{z}}_n  & = U \eee{-\ii \alpha_n}+ \mu \conj{w}(z_n)+ V_n, \\[1ex]
  \dot{\alpha}_n  & =  \real{\nu_1 \frac{d \conj{w}}{dz} \ii \eee{2 \ii \alpha_n}+\nu_2 \conj{w} \ii  \eee{\ii \alpha_n}}.
  \end{split}
\end{equation}
Here,  $z_n$ and $\alpha_n$ denote the position and orientation of each swimmer ($n=1,\ldots, N$), Re denotes the real part of the expression in bracket, whereas $\mu$, $\nu_1$ and $\nu_2$ are non-dimensional translational and rotational mobility coefficients whose values depend on the translational mobility coefficients  $\mu_{head}$ and $\mu_{tail}$ at the swimmer's head  and tail~\cite{supp}.
A swimmer can have different mobility coefficients  $\mu_{head}$ and $\mu_{tail}$, with $\mu_{tail}$ being a decreasing function of $R_{tail}$. Given that $\mu_{head}$ is constant, the value of $\mu_{head}-\mu_{tail}$ can thus change sign from positive to negative by decreasing the level of flagellar activity, and vice versa.   This signed difference  dictates the signed value of $\nu_2$ and therefore how a swimmer reorients in response  to the local flow.  Vigorous flagellar activity 
for which $\mu_{head}-\mu_{tail}>0$ (i.e., $\nu_2>0$) causes swimmers to reorient in the direction of the local flow whereas swimmers with weakly-beating flagella  for which $\mu_{head}-\mu_{tail}<0$ (i.e.,  $\nu_2<0$) reorient in the opposite direction to the local flow. The swimmers also reorient in response to the flow gradient as indicated by the $\nu_1$-term in~\eqref{eq:formulation:eom} consistently with the classical  Jeffery's orbit~\cite{jeffery:prsa1922a}.


To close the model in~\eqref{eq:formulation:eom}, we need to evaluate the velocity field $w(z)$ induced  by $N$ potential dipoles in a doubly-periodic domain, which involves the evaluation of conditionally-convergent, doubly-infinite sums of terms that decay as $1/|z|^2$. In this letter, we present a novel, closed-form solution for this doubly-periodic system. We distinguish our exact analytical solution 
from the approximate numerical solution in~\cite{lefauve:pre2014a}.  
We showed in~\cite{tsang:jnls2013a} that the velocity field associated with a system of \emph{finite dipoles}  in a doubly-periodic domain can be expressed in terms of the Weierstrass zeta-function and, in~\cite{kanso:fdr2014a}, we derived a point dipole model  that is consistent with both the finite dipole system and the model in~\eqref{eq:formulation:eom}. Building upon these results, we get, after some straightforward but tedious manipulations, that the velocity field induced by $N$ potential dipoles located at $z_n$ with orientation $\alpha_n$, $n=1,\ldots, N$, in a doubly-periodic domain can be written in terms of the Weierstrass elliptic function
as follows:
\begin{equation}
	\label{eq:formulation:velocityDipolePeriodic}
	 \conj{w} = \sum_{n=1}^N\sigma_n \rho( z-z_n;\omega_1,\omega_2)\eee{\ii \alpha_n}.
\end{equation}
Here, $\sigma_n$ is the strength of the potential dipole associated with the $n^{th}$ swimmer. 
The Weierstrass elliptic function $\rho(z)$ is given by $\rho\left(z;\omega_{1},\omega_{2}\right)=\frac{1}{z^2}+\sum_{k,l} \left(\frac{1}{(z-\Omega_{kl})^2} -\frac{1}{\Omega_{kl}^{2}}\right)$, with $\Omega_{kl}=2k\omega_{1}+2l\omega_{2}$,  $k,l\in \mathbb{Z} \!-\!\{0\}$,
and $\omega_{1}$ and  $\omega_{2}$ being the half-periods of the doubly-periodic domain. This function has infinite numbers of double pole singularities located at positions of $z=0$ and $z=\Omega_{kl}$, corresponding to the $1/|z|^2$ singularities induced by the potential dipoles.

In addition to the hydrodynamic coupling, we account for steric interactions in~\eqref{eq:formulation:eom} using a collision avoidance mechanism $V_n$
based on the repulsive part of the Leonard-Jones potential.
These near field interactions decay rapidly outside a small excluded area centered around $z_n$. Their rapid decay ensures that the order of the far-field hydrodynamic interactions is preserved.

\begin{figure}[!b]
\centerline{\includegraphics[width=0.5\textwidth]{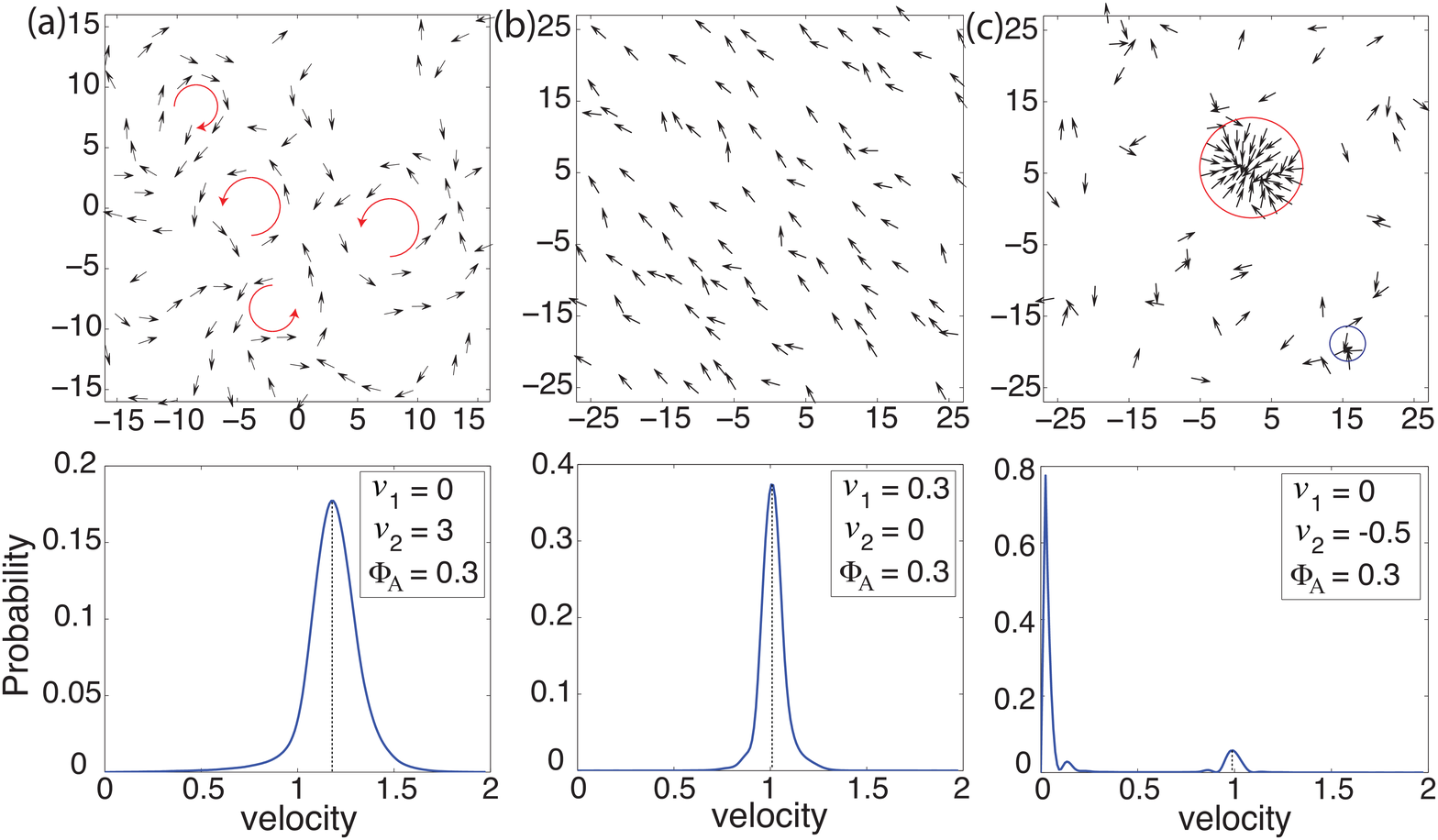}}
\caption[]{(a) Swirling motion in vigorously-beating flagellated swimmers; 
(b) orientational oder; (c) aggregation in weakly-beating flagellated swimmers.
Probability distributions of velocity of the swimmers are depicted in the lower row.}
	\label{fig:emergence}
\end{figure}

\begin{figure}
\centerline{\includegraphics[width=0.5\textwidth]{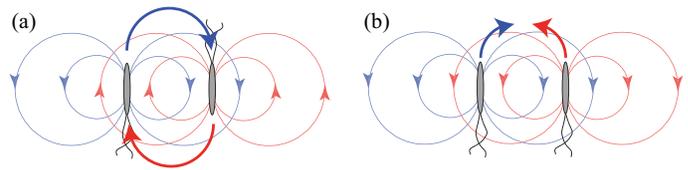}}
 \caption[]{(a) Swimmers with vigorously-beating flagella orient with local flow and thus tend to chase each other; (b) Swimmers with weakly-beating flagella orient in the opposite direction to local flow and thus tend to aggregate together.}
	\label{fig:asymmetricschematic}
\end{figure}

We focus on the evolution of populations of micro-swimmers that are initially randomly oriented but spatially homogenous. We normalize $U$ and $\sigma_n$ to 1 and set $\mu = 0.5$, $N = 100$. 
We vary $\nu_1$, $\nu_2$ and the area fraction $\Phi_A= NA/A_0$, where $A$ is the area of the micro-swimmer, and $A_0 = L^2$ is the size of the doubly-periodic square domain. 
We perform Monte-Carlo type simulations in the sense that, for each set of parameters ($\nu_1$, $\nu_2$, $\Phi_A$), we run multiple trials corresponding to different sets of initial conditions~\cite{supp}. 
We observe the emergence of three distinct types of global structures:  swirling behavior for vigorously-beating flagella, orientational order for a narrow range of parameter values around $\nu_2 = 0$, and aggregation or clustering for weakly-beating flagella. Representative simulations are shown in Fig.~\ref{fig:emergence}. 
The swirling-like motion is characterized by a velocity distribution function with a mean value higher than the speed of the individual swimmer (Fig.~\ref{fig:emergence}(a)). 
This collective behavior where vortex-like structures emerge, break, and rearrange elsewhere is reminiscent to the bacterial turbulence observed in numerous experiments~\cite{mendelson:jbact1999a,dombrowski:prl2004a,cisneros:ef2007a,sokolov:prl2007a,sokolov:pre2009a,darnton:bj2010a,dunkel:prl2013a}. 
 The associated increase in swimming speed was also observed experimentally.
In the context of the dipole model, the increase in speed can be explained as follows.  Swimmers with vigorously-beating flagella  tend to ``tail-gate" each other as a  result of them aligning with the local flow field. When a swimmer is close to another swimmer, it will orient towards and travel along the streamlines of the potential dipole created by the nearby swimmer as depicted schematically in Fig.~\ref{fig:asymmetricschematic}(a). As the swimmers align and form a chain-like structure, they create a flow field that helps their forward motion, thus increasing the swimmers' velocities, as evidenced from the probability distribution function in Fig.~\ref{fig:emergence}(a).

As the flagellar activity decreases (by decreasing the value of $\nu_2$), a transitional behavior that does not exhibit a clear pattern is observed before a global orientational order develops,
see Fig.~\ref{fig:emergence}(b). 
The development of orientational order happens at a much longer time scale than the swirling dynamics  and is a result of the dipoles reorienting with the local velocity gradient (non-zero $\nu_1$), which is ignored in~\cite{brotto:prl2013a, lefauve:pre2014a}. In this case, the velocity distribution function is Gaussian centered at $U=1$ (Fig.~\ref{fig:emergence}(b)), implying that this collective mode has no advantage at the population level in terms of increased swimming speed.

When we further decrease the flagellar activity, the swimmers begin to aggregate and cluster, see Fig.~\ref{fig:emergence}(c). The clustering behavior of swimmers with weak flagellar activity can be explained by recalling that such swimmers reorient in the opposite direction of the local flow field. Therefore, a swimmer tends to travel in the opposite direction to the streamlines created by a nearby swimmer, see Fig.~\ref{fig:asymmetricschematic}(b), which leads to aggregation. The collective aggregation of many swimmers takes place at a very short time scale and leads to the formation of clusters. Some clusters are unstable and break readily (blue circle in Fig.~\ref{fig:emergence}(c)).  However, long-lived stable clusters also form (red circle in Fig.~\ref{fig:emergence}(c)). The stable cluster depicted in Fig.~\ref{fig:emergence}(c)  slows down considerably as more swimmers join the cluster. The velocity distribution function has a strong peak around zero velocity because most swimmers are attached to the stable cluster, with the remaining swimmers moving at unit speed.

\begin{figure}
\includegraphics[width=0.5\textwidth]{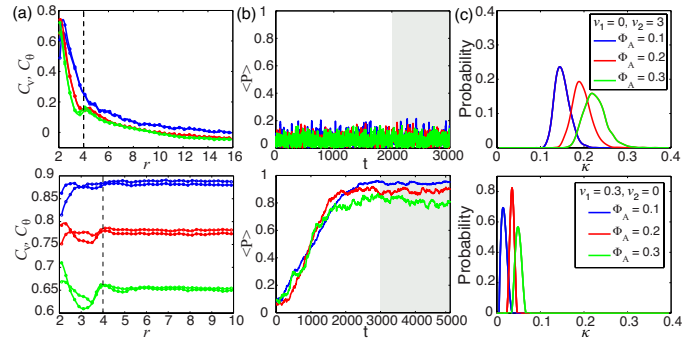}
 \caption[]{Polar order parameter $\langle P \rangle$, velocity and angular correlations $C_{\rm v}$, $C_\theta$, probability distribution of rotational activity parameter $\kappa$, all computed based on data corresponding to the shaded time range.}
	\label{fig:statistics}
\end{figure}
\begin{figure}
\centerline{\includegraphics[width=0.475\textwidth]{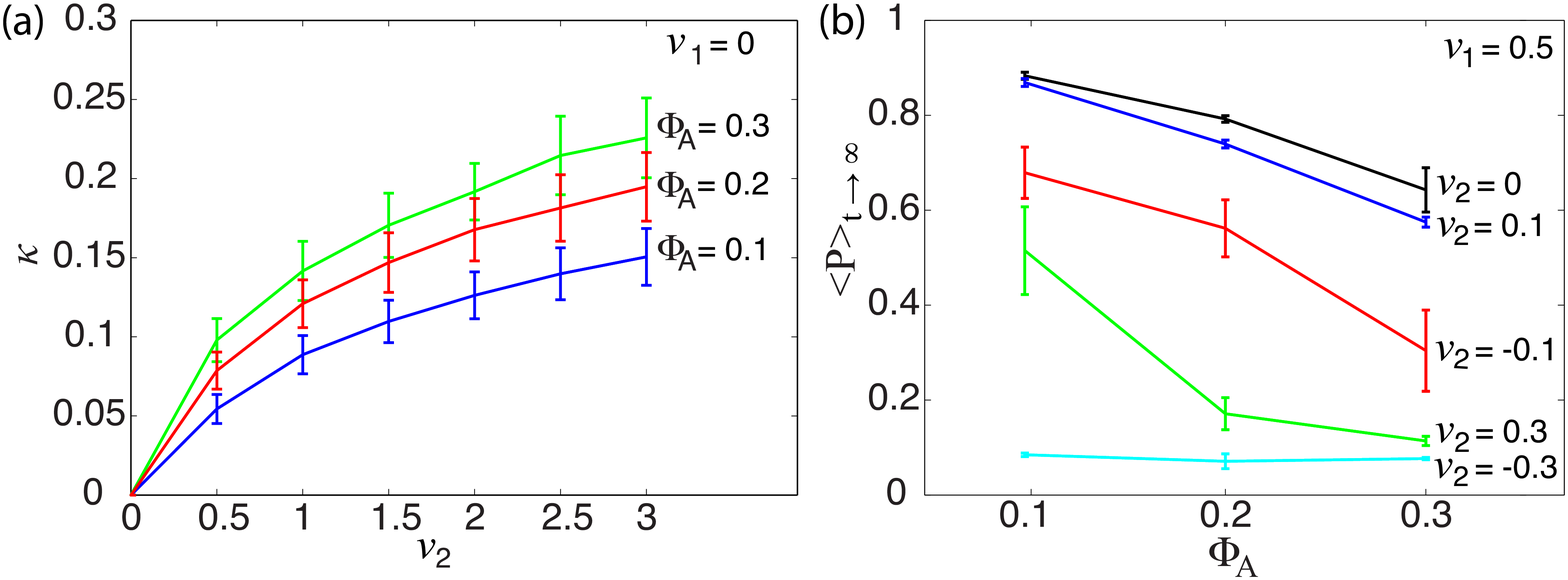}}
\caption[]{(a) The time-averaged mean and standard deviation of $\kappa$ and (b) the sample-averaged mean and standard deviation of long-time developed $\langle P \rangle$ over various initial conditions.
}	\label{fig:kappavariation}
\end{figure}
\begin{figure}
\centerline{\includegraphics[width=0.5\textwidth]{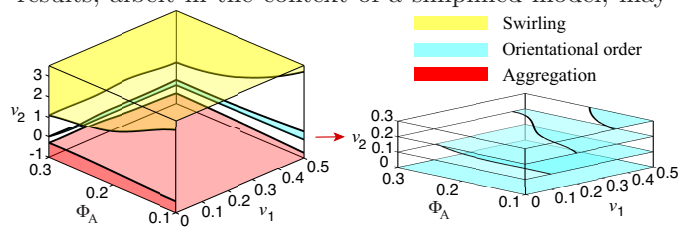}}
 \caption[]{Phase diagrams showing the three different collective modes, swirling, orientational order and aggregation, as a function of $\nu_1$, $\nu_2$ and $\Phi_A$. Regions of transitional behavior for which no clear pattern are identified are left in white.}
	\label{fig:phasediagram}
\end{figure}

We use a number of statistical measures to assess the observed global structures.  In addition to the velocity distribution function shown in Fig.~\ref{fig:emergence},  we compute the velocity and angular correlation functions $C_{\rm v}$ and $C_{\theta}$  as a measure of the spatial range for which the velocity and orientation of a swimmer are coordinated with those of its neighbors. Local correlations are observed for the swirling-type motion while global correlations are seen when orientational order is developed (Fig.~\ref{fig:statistics}(a)). We also compute the polar order parameter $\langle P(t) \rangle = \frac{1}{N} \abs{\sum_{n=1}^N \eee{\ii \alpha_n(t)}}$
to assess the degree of global order in the swimmers population (Fig.~\ref{fig:statistics}(b)). However, these statistical functions do not distinguish between the swirling behavior 
and the transitional behavior between the three global modes reported here. Therefore, we introduce a rotational activity parameter $\kappa(t) = \frac{1}{N} \sum_{n=1}^N {|\dot{\alpha}_n(t) |}/{|\dot{z}_n (t) |}$ that measures the average change in orientation weighted by the traveling distance (Fig.~\ref{fig:statistics}(c)). The values of $\kappa$ exhibit a continuous transition as $\nu_2$ decreases as shown in Fig.~\ref{fig:kappavariation}(a). Fig.~\ref{fig:kappavariation} also shows that the general dependence of $\kappa$ and  long-time developed $\langle P \rangle$ on $\nu_2$ and $\Phi_A$ is  insensitive to variations in initial conditions. We set threshold values for $\kappa$ and $\langle P \rangle$, and use the mean of the velocity distribution function to distinguish between the three global modes: swirling,  orientational order, and aggregation. These behaviors are mapped  onto the 3D  parameter space ($\nu_1$, $\nu_2$, $\Phi_A$) in the phase diagram depicted in Fig.~\ref{fig:phasediagram}. Note that the orientational order mode arises in a narrow parameter range where $\nu_2$ is small and the hydrodynamic interactions are dominated by the weaker gradient term in~\eqref{eq:formulation:eom},  which explains why it develops at a much slower time scale. This mode is therefore expected to be less robust to rotational Brownian noise than the swirling and clustering modes. 

 
To conclude, we note that, whereas our finite-sized system is fully-nonlinear and non-dilute, the $(\nu_2, \phi_A)$-slice of Fig.~\ref{fig:phasediagram} may be interpreted as a rough analog to the 2D  linear stability diagram obtained in the kinetic model of~\citep[Fig. 2]{brotto:prl2013a}, where $\nu_1$ was considered identically zero. The linear instabilities reported in the kinetic model do not reveal the nature of the emergent collective modes (swirling versus clustering) nor their physical implications. Here, we showed, for the first time, that transitions from swirling to clustering and aggregation, and vice-versa, can be induced by an interplay between the flagellar activity and the hydrodynamic interactions. These results, albeit in the context of a simplified model, may have profound implications  on understanding biofilm formation, which often occurs in response to overcrowding or nutrient depletion that cause decreased flagellar activity. They suggest that the interplay between hydrodynamics and flagellar activity may serve as a physical mechanism for initiating the biofilm state by aggregating the bacteria, which would concentrate their chemical secretions in preparation of the biofilm formation.









\bibliography{reference}

\end{document}